\documentclass[10pt,twocolumn,english,aps,prl,showpacs,superscriptaddress]{revtex4-1}
\usepackage[latin9]{inputenc}
\usepackage{color}
\usepackage{babel}
\usepackage{units}
\usepackage{textcomp}
\usepackage{amsmath}
\usepackage{amssymb}
\usepackage{stmaryrd}
\usepackage{graphicx}
\usepackage[unicode=true,pdfusetitle,
 bookmarks=true,bookmarksnumbered=false,bookmarksopen=false,
 breaklinks=false,pdfborder={0 0 0},backref=false,colorlinks=true]
 {hyperref}

\makeatletter

\@ifundefined{textcolor}{}
{%
 \definecolor{BLACK}{gray}{0}
 \definecolor{WHITE}{gray}{1}
 \definecolor{RED}{rgb}{1,0,0}
 \definecolor{GREEN}{rgb}{0,1,0}
 \definecolor{BLUE}{rgb}{0,0,1}
 \definecolor{CYAN}{cmyk}{1,0,0,0}
 \definecolor{MAGENTA}{cmyk}{0,1,0,0}
 \definecolor{YELLOW}{cmyk}{0,0,1,0}
}

\usepackage[usenames,dvipsnames]{xcolor}
\usepackage{epstopdf}
\newcommand*{\mathcolor}{}
\def\mathcolor#1#{\mathcoloraux{#1}}
\newcommand*{\mathcoloraux}[3]{%
  \protect\leavevmode
  \begingroup
    \color#1{#2}#3%
  \endgroup
}

\makeatother

\begin{document}

%=========== TITLE
\title{Berry phase of  light Bragg-reflected by chiral liquid crystal media}

%=========== AUTHORS
%--- R. Barboza
\author{Raouf Barboza}
\email{raouf.barboza@ing.uchile.cl}
\affiliation{Departamento de F\'isica, FCFM, Universidad de Chile, Casilla 487-3 Santiago,Chile}
%--- U. Bortolozzo
\author{Umberto Bortolozzo}
%\email{umberto.bortolozzo@inln.cnrs.fr}
\affiliation{INLN, Universit\'e de Nice-Sophia Antipolis, CNRS, 1361 Route des Lucioles, 06560 Valbonne, France}
%--- M.G. Clerc
\author{Marcel G. Clerc}
%\email{marcel@dfi.uchile.cl}
\affiliation{Departamento de F\'isica, FCFM, Universidad de Chile, Casilla 487-3 Santiago,Chile}
%--- S. Residori
\author {Stefania Residori}
%\email{stefania.residori@inln.cnrs.fr}
\affiliation{INLN, Universit\'e de Nice-Sophia Antipolis, CNRS, 1361 Route des Lucioles, 06560 Valbonne, France}

%=========== ABSTRACT
\begin{abstract}
Berry phase is revealed for circularly polarized light when it is 
Bragg-reflected by a chiral liquid crystal medium of the same handedness. 
By using a chiral nematic layer we demonstrate that if the input plane of the 
layer is rotated with respect to a fixed reference frame, then, a geometric 
phase effect occurs for the circularly polarized light reflected by the 
periodic helical structure of the medium. Theory and numerical simulations are 
supported by an experimental observation, disclosing novel applications in 
the field of optical manipulation and fundamental optical phenomena.
\end{abstract}

%=========== PACS
\maketitle
\date{\today}

Berry phase is a phenomenon well known to occur when dealing with cyclic adiabatic transformations \cite{Berry45}. Introduced by Berry for quantum 
systems undergoing a cyclic evolution under the action of a time-dependent Hamiltonian \cite{Berry45}, the concept has been generalized to the accumulation of a 
geometrical phase arising from topological aspects in different quantum and classical systems, among the most known the Aharonov-Bohm effect and the precession 
of the oscillation plane of a Foucault pendulum (see, e.g., \cite{Zela} for a review). Berry phase for light is associated to geometric manipulation of the polarization and/or the 
direction of propagation, see \cite{Bliokh_PRL2008} and reference therein. Geometric phase due to the polarization state manipulation was discovered by the pioneering work of Pancharatnam, 
who highlight it as an intrinsic properties of polarized light \cite{Pancharatnam}. Indeed, when the polarization of a beam describes  a closed-loop on the Poincar\'e sphere, the 
final state differs from the initial one by a geometric phase, Pancharatnam-Berry (PB) phase, proportional to the area enclosed by the loop. Based on this feature, 
space-variant waveplates have been specifically designed to realize PB phase optical 
elements with subwavelenght gratings, such as polarization sensitive 
diffraction gratings \cite{Hasman_0L2002} 
and spiral phase elements \cite{Hasman_OL2002_helical}, for laser radiation 
at $10.6$ $\mu m$ wavelength. 
PB phase optical elements in the visible spectral domain have, then, been 
achieved by using nematic liquid crystals with patterned alignment, such as 
q-plates that generate helical modes of visible light \cite{Marrucci_PRL2006,Tabiryan_OPEX2013}. 
Based on nematics, other PB phase optical devices, such as switchable lenses, 
beam splitters, and holographic elements, have been proposed \cite{Marrucci_APL2006} 
and, more recently, a diffractive lens with convergent/divergent behavior depending on the 
left/right handedness of the circular polarization of the input light has been realized \cite{Escuti_Optica2015,Chigrinov_SID2015}. 

\begin{figure}[b]
\centering
\includegraphics[width=\columnwidth]{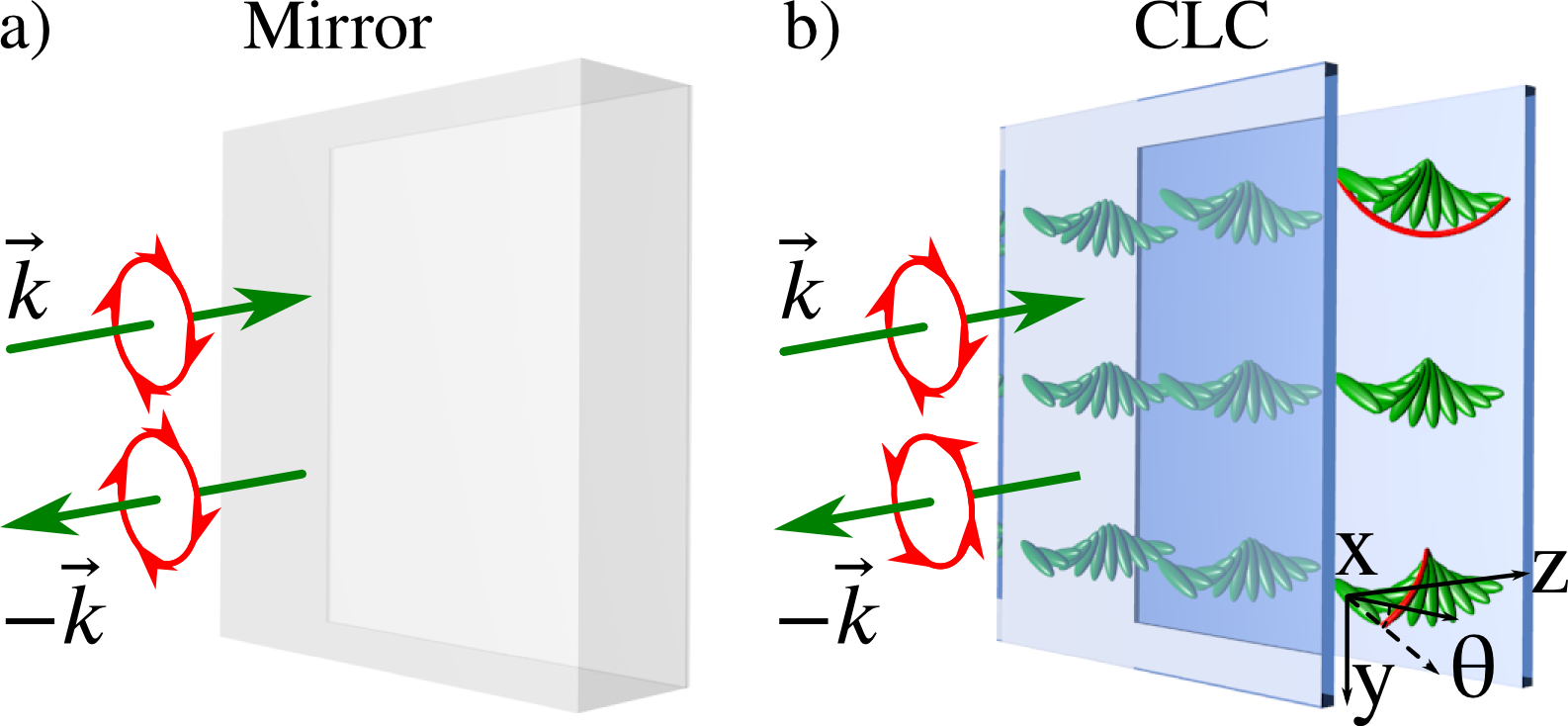}
\caption{(color online). Reflection from (a) a conventional mirror and (b) a chiral liquid crystal  (CLC) medium, the red arches depict the cholesteric helix; while the mirror reverses the polarization of the reflected light, a circularly polarized beam 
matching the handedness of the chiral helix maintains the same polarization when reflected.}
\label{mirror}
\end{figure}

Here, we demonstrate that a Berry phase effect occurs for 
circularly polarized light when it is Bragg-reflected by a chiral liquid crystal 
medium. The optical wavelength and polarization handedness of the incident 
beam match, respectively, the Bragg wavelength and the handedness of the helical structure in the chiral medium.
The geometric phase is introduced by the symmetry breaking in spin/momentum (handedness/direction) space  and can be revealed by a space-variant optical axis of the 
uniaxial chiral layer. At the purpose of revealing the effect we employ a chiral 
nematic liquid crystal (CLC) cell
with appropriately designed anchoring conditions at the entrance plane.
Unlike conventional reflective elements, the meaningful component of the phase, here, is not introduced by 
optical path differences, but it results from the geometrical phase accumulated 
at the input plane of the CLC cell. Indeed, if at the input plane the CLC layer is 
rotated with respect to a fixed reference frame, then, a geometric phase effect 
occurs during the reflection. By exploiting this property, and by introducing 
patterns of molecular anchoring at the entrance plane of the CLC cell, 
reflectors performing different wavefront shaping can be designed and realized.

A qualitative comparison between a conventional mirror and a CLC medium is sketched 
in Fig.~\ref{mirror}. At variance with the mirror (Fig.~\ref{mirror}a), that reverses both the 
propagation direction and the polarization of the reflected light, the CLC cell (Fig.~\ref{mirror}b) 
maintains the same polarization of the reflected beam. Indeed, CLC are characterized 
by a self-assembled helical structure of their molecular arrangement and selective Bragg reflection: when light is circularly polarized with the same handedness of the chiral helix, 
the periodic helical structure leads to Bragg reflection at normal incidence whenever the optical wavelength 
$\lambda$ satisfies the condition 
$n_o P <\lambda <n_e P$, where $P$ is the chiral pitch and $\{n_o, n_e\}$ 
are the ordinary and extraordinary refractive indices, respectively 
\cite{deGennes:1995}. The circularly polarized component with opposite 
handedness is transmitted through the chiral layer. 
These properties combined with the ability to change their pitch with electric 
or magnetic fields \cite{Yeh:1999}, with temperature \cite{Tzeng} or light \cite{Dong_OPEX2012}, 
make CLC attractive for applications as dynamically controlled reflection/transmssion 
filters \cite{Mitov_Nature2006}, tunable lasers \cite{Coles_NAtPhot2010} and 
photoswitchable bandgaps \cite{Tabiryan_OPEX2010}. Here, we show that a 
Berry phase is always associated with the Bragg reflection at normal incidence, which allows adding 
novel functionalities to CLC layers, such as the ability to arbitrarily shaping 
the reflected wavefront.

\section{Detailed explanation}
A detailed explanation of such geometric phase can be derived by considering a perfectly plane CLC layer with the director at its front face orientated at an angle $\theta$ with respect to 
the $x$-axis of a fixed reference frame
$(x,y,z)$, $\hat{z}$ coinciding with the axis of the chiral helix and with the propagation direction of the incident beam. 
We can now introduce a local coordinate system $(x',y',z)$, such that $\hat{x}'$ is parallel to the 
molecular director at the input face of the CLC cell, that is, $\hat{x}'=\hat{x}\cos\theta+\hat{y}\sin\theta$, and  $\hat{y}'=-\hat{x}\sin\theta+\hat{y}\cos\theta$.
In the fixed reference frame a $\sigma$-handed circular polarization writes as $\hat{e}_\sigma=(\hat{x}+i\sigma\hat{y}) /\sqrt{2}$, which, in the local coordinate system, transforms to 
$\hat{e}'_\sigma=(\hat{x}'+i\sigma\hat{y}')/\sqrt{2}=e^{-i\sigma\theta} \hat{e}_\sigma$. 
Consider now a circularly polarized incident plane wave $\vec{E}_i=E_i(0)e^{i(kz-\omega t)}\hat{e}_\sigma$, with its wavelength inside the reflective bandgap of the CLC, its  
polarization handedness matching that of the chiral helix and at normal incidence. The incident wave is totally reflected by the CLC layer and its handedness is conserved after the reflection. 
In the fixed reference frame the reflected wave can, thus, be written as $\vec{E_r}=E_r(0)e^{-i(kz+\omega t)}\hat{e}_{-\sigma}$. 
If the axis of the CLC layer is chosen to coincide with $\hat x$ ($\theta=0$), then, we can apply the well known relation
$E_r(0)=rE_i(0)$, with $r$ the reflection coefficient \cite{Yeh:1999}. However, if the CLC is rotated by an angle $\theta$ with respect to $\hat x$, then, the relation between 
the amplitude of the incident and the reflected electric field must be considered in the local coordinate system and the phase  arising from the coordinate transform from the 
local to the fixed reference must be taken into account, they do not  necessarily cancel out. 
Therefore, the incident and the reflected beam, respectively, $\vec{E}_i=E_i\hat{e}_\sigma$ and $\vec{E}_r=E_r\hat{e}_{-\sigma}$ in the local coordinate system write as $\vec{E}
_i=E'_i\hat{e}'_\sigma=E'_ie^{-i\sigma\theta}\hat{e}_\sigma$
and $\vec{E}_r=E'_r\hat{e}'_{-\sigma}=E'_re^{i\sigma\theta}\hat{e}_{-\sigma}$, respectively. The explicit dependence on $z$ and $t$ are dropped in sake of clarity.
As a consequence, the  reflection coefficient is given  by $r= E_r(0)/E_i(0)=e^{2i\sigma\theta} E'_r(0)/E'_i(0)$, which can also be written as 
$r=r' \, e^{i \Phi_{B}}$ where $\Phi_{B}= 2\sigma\theta$ is the Berry phase, $r'$ the reflection coefficient in the local coordinate and does not depend on $\theta$. Note that  
$\Phi_{B}$ has a pure geometric origin, that is, it does not originate from an extra optical path but from a purely geometrical transformation, namely, the flipping of the propagation 
direction while maintaining the same helicity of the circular polarization, and the rigid rotation of the CLC layer. 
Based on the same analysis it is easy to show that the net geometric phase effect is cancelled for the transmitted component (case of thin layers). 
Indeed, in such case the amplitude of the transmitted wave can be written as $\vec{E}_t=E_te^{i(kz-\omega t)}\hat{e}_\sigma$, which transforms 
in the local coordinate system as the incindent beam, i.e. $\vec{E}_t=E'_t\hat{e}'_\sigma=E'_t e^{i\sigma\theta}\hat{e}_\sigma$, and, correspondingly, the transmission coefficient 
$t=E_t(L)/E_i(0)=E'_t(L)/E'_i(0)$, thus $t=t'$, $t'$ being the transmission coefficient in the local reference, independent from $\theta$.
\begin{figure} [t]
\centering
\includegraphics[width=0.8\columnwidth]{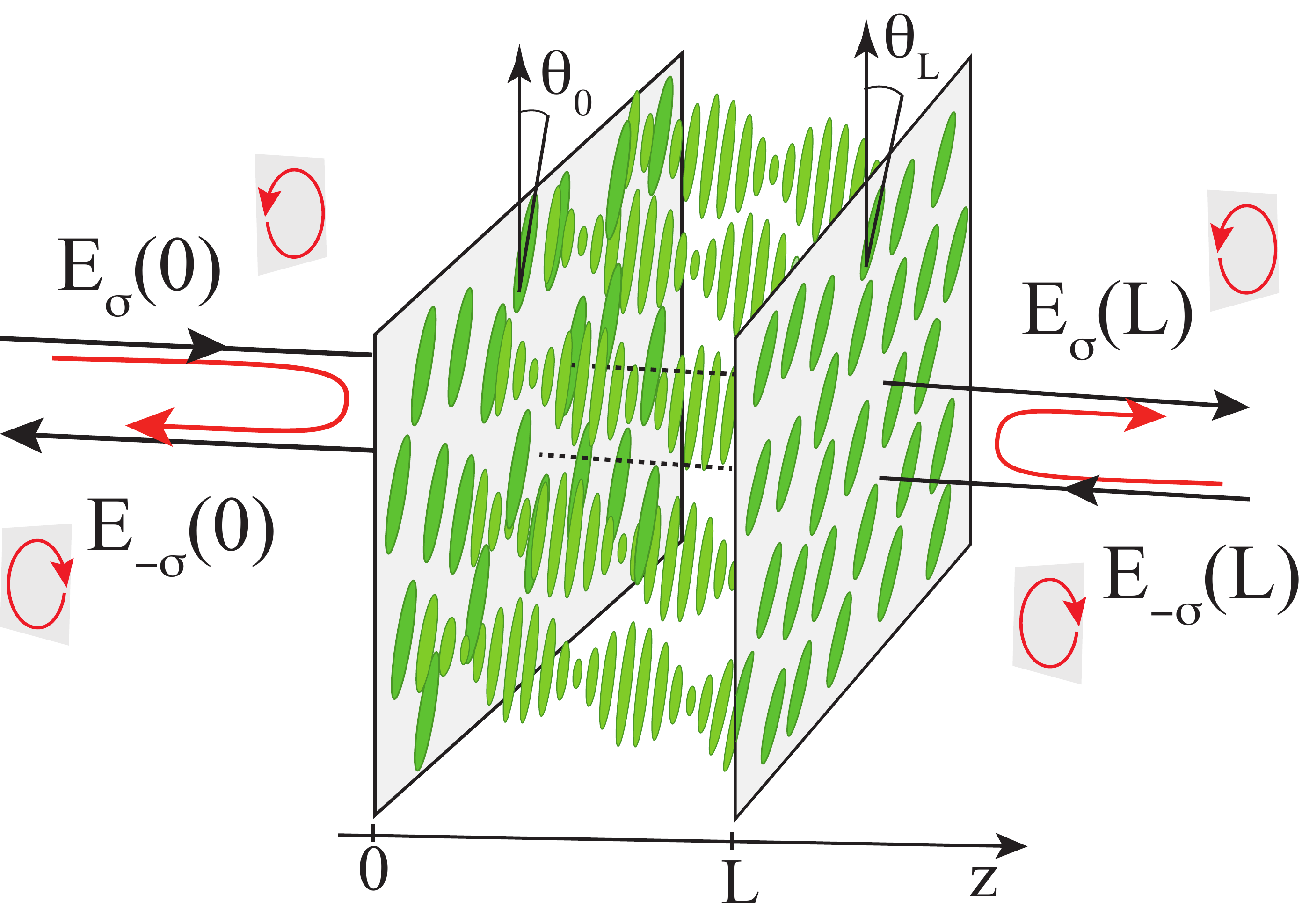}
\caption{(color online). Modes of light propagation in the CLC layer. Only the Bragg regime is considered; the (red) curved arrows indicate the Bragg-reflected beam to which it is associated the Berry phase.}
\label{scattering}
\end{figure}

The dependence of the reflection coefficient on the molecular angle $\theta$ at the input plane of the CLC layer can be derived analytically from the light propagation equations and by 
using the coupled mode theory. Such treatment can be found in the Supplemental Material \cite{Supplementary}. 
As a result, all the possible cases can be simply summarized in a scattering matrix, which reads as 
\begin{equation}
\begin{bmatrix}E_{\sigma}(L)\\E_{-\sigma}(0)\end{bmatrix}=\begin{bmatrix}t & r'\, e^{-2i\sigma\theta_L}\\ r' \, e^{2i\sigma\theta_0} & t\end{bmatrix}\begin{bmatrix}E_{\sigma}(0)\\E_{-\sigma}(L)
\end{bmatrix},
\end{equation}
where $E_{\sigma}(0)$ is a $\sigma$-handed circularly polarized wave incident 
on the front face of the CLC layer and $E_{-\sigma}(L)$ is a circularly polarized 
wave with the same handedness but propagating in the opposite direction and 
incident on the end face of the CLC layer. The output fields $E_{\sigma}(0)$ 
and $E_{\sigma}(L)$ are built each by two contributions: the beam transmitted 
through the CLC layer from the other end of the cell and the beam propagating 
from the same side and Bragg-reflected by the periodic helix in the CLC layer. 
From the scattering matrix terms it can be seen that the reflection coefficients 
are accompanied by a geometric phase, respectively ${\Phi_{B}}_0=2\sigma \theta_0$ 
and ${\Phi_{B}}_L=-2\sigma \theta_L$, where the angles $\theta_0$ and $\theta_L$ 
denote the anchoring direction of the molecules at the entrance, respectively, 
exit plane of the CLC layer. The derived scattering matrix is similar to those used to describe 
transmissive and reflective anistropic uniaxial geometric-phase element,  see \cite{Bliokh:NatPhot:2015} and 
references therein. Rather than linking the reflected light or the transmitted light to the incident field, it connects circularly polarized components  of reflected and transmited waves to the forward an the backward propagating circularly polarized incident fields; the handness considered in that of the cholesteric cell. Morever, unlike for the untwisted structures, the phase of the off-diagonal terms (that carry the information about the geometric phase) can be made independent with dissimilar anchoring angle at the faces of the geometric phase element. A scheme representing the scattering processes is displayed in Fig.~\ref{scattering}. 

\section{Numerics}
\begin{figure}[b]
\centering
\includegraphics[width=0.7\columnwidth]{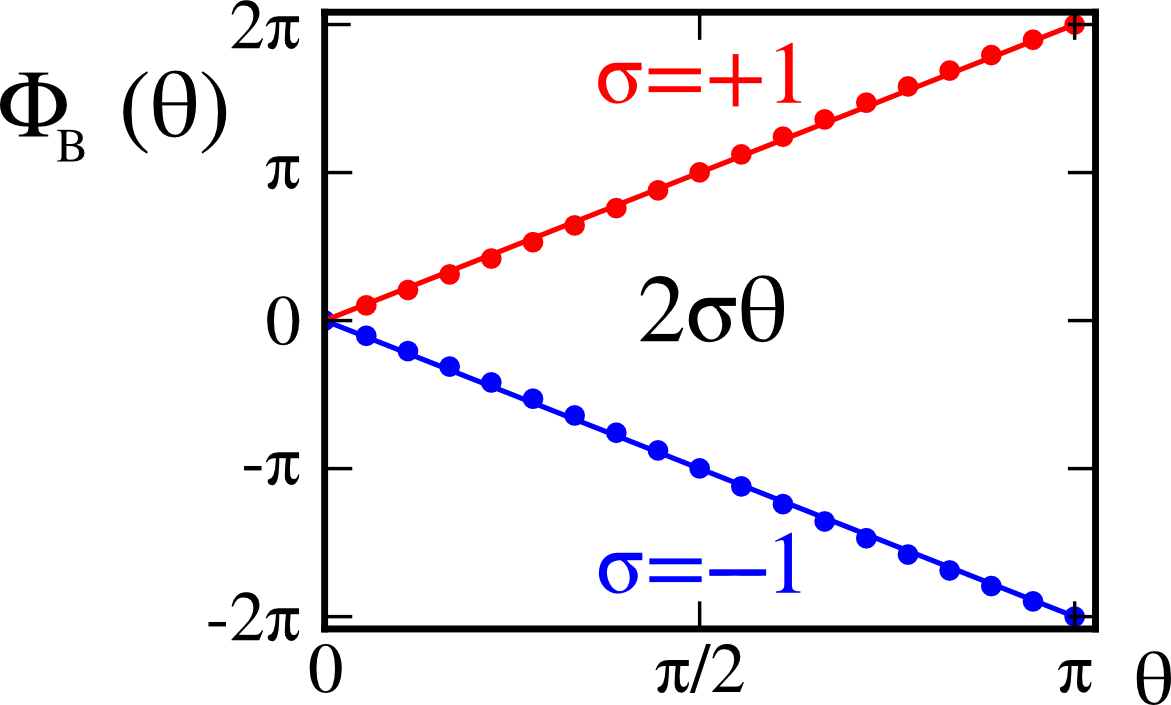}
\caption{ Numerically calculated Berry phase versus the angle $\theta$ of the molecular alignment at the input plane of the CLC cell, the results 
(dots) match their corresponding (lines) analytical prediction $\Phi_{B}(\theta)=2 \sigma \theta$, $\sigma=\pm1$ being the handedness of cholesteric 
helix and the input polarization.}
\label{FDTD}
\end{figure}

Numerical simulations were performed with the finite-difference time-domain (FDTD) method \cite{Taflove:2005}, using a freely available software package \cite{MEEP}.
We have calculated the geometric phase, taking phase of the 
reflection coefficient of a left/right handed ($\sigma=\pm1$) circularly 
polarized beam by left/right handed helix \cite{Supplementary}. 
The results are shown in Fig. \ref{FDTD}, where $\Phi_{B}$ is plotted 
versus $\theta$ the anchoring angle at the input of the cell. The right handed case and the left handed case were considered. The geometric phase varies linearly and the slope depends on 
the handedness of the polarization, $\Phi_{B}(\theta)=2 \sigma \theta$ can be clearly verified.

\section{Experiment}
\begin{figure}[t!]
\centering
\includegraphics[width=\columnwidth]{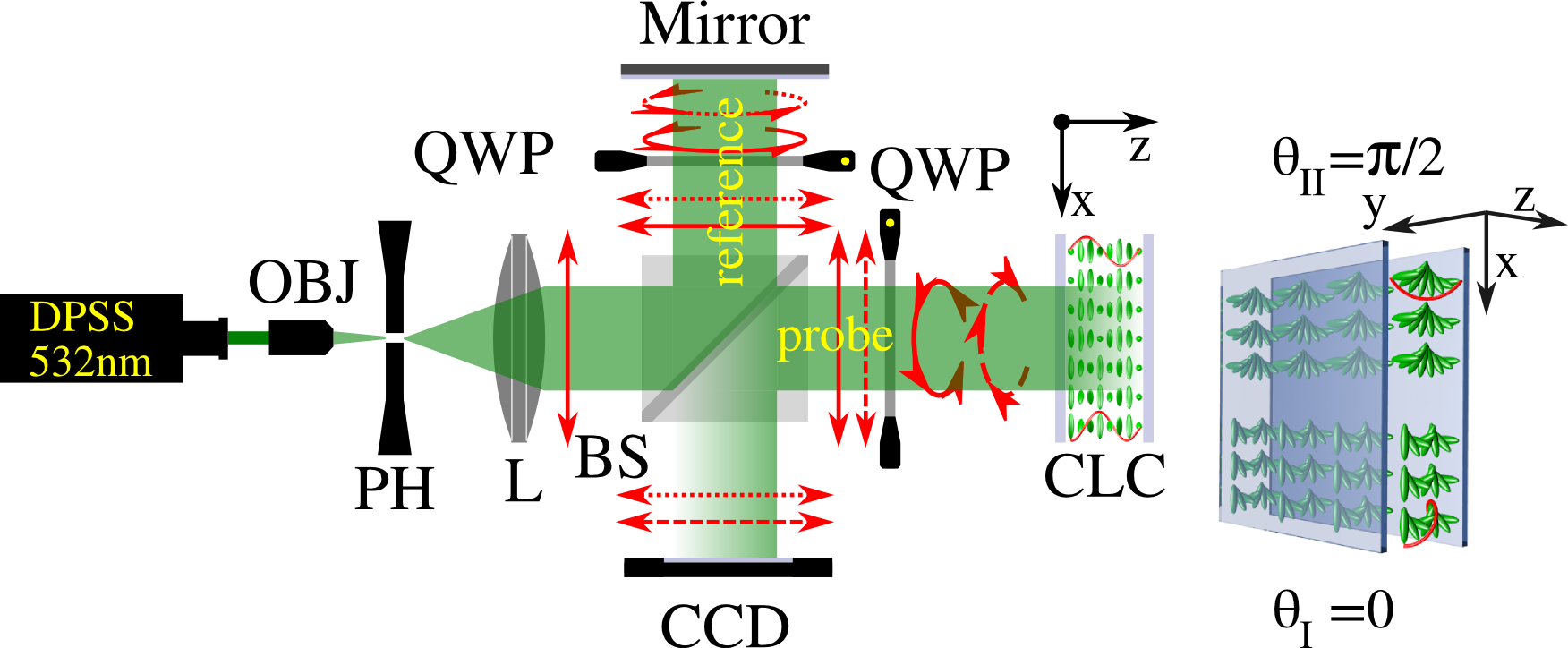}
\caption{(color online). Experimental setup: a diode pumped solid state (DPSS) 
laser at $532$ $nm$ is divided into a reference and a probe beam by the 
beam-splitter (BS), then, recombined in a Michelson type interferometer onto a CCD 
camera; while the reference is reflected by an ordinary mirror the probe beam is 
retro-reflected by a CLC cell which has two regions of planar alignment 
in its entrance plane (anchoring angles $\theta_I=0$ and $\theta_{II}=\pi/2$), as 
shown in the inset. The red helical arches depict the cholesteric helix for a 
better clarity on the handedness. OBJ: objective; PH: pinhole; L: lens; QWP: 
quarter-wave plate.
}
\label{setup}
\end{figure}

To reveal experimentally the effect we have setup a Michelson type 
interferometer in which one of the mirrors is substituted by a CLC cell, as 
depicted in Fig\ref{setup}. The cell is filled with a CLC mixture composed of 
the nematic E7 ($n_e=1.7472$ and $n_o=1.5217$ at $589.3$ $nm$ at $20^\circ$) 
doped with the chiral agent CB15 ($41\%$ in weight). The chiral mixture is 
injected between two planar rubbed polyvinyl alcohol(PVA)-coated glass plates 
with $L=9$ $\mu m$ thick spacers. The front plane of the cell has been 
rubbed mechanically in two step. After rubbing the first half of the glass 
plate, the second half is rubbed in a direction orthogonal to the previous one.
This provides us with with two regions of orthogonal planar  alignment, the 
anchoring angle being $\theta_I=0$ and $\theta_{II}=\pi/2$ in each region, 
respectively. The helical structure of the chiral mixture is perpendicular to 
the confining walls and has a pitch $P$ such that $\sim 50$ half-pitches are 
contained along the cell thickness. Note that the cell can be considered thick 
insofar as just a few pitches of the helix close to the entrance plane are 
enough to provide the total construction of the Bragg-reflected beam. 
%------------------ 
The incident beam is circularly polarized with the same handedness of the chiral 
layer and overlaps the two regions of orthogonal anchoring. Because of the 
orthogonal anchoring conditions, the two regions of the CLC cell induce 
a geometric phase shift on the reflected wavefront. 
The phase shift is revealed by interfering the probe with the reference wave, which 
is reflected by an ordinary mirror. A quarter-wave plate is used to convert the 
probe to a circularly polarized wave with the same handedness as the CLC helix. 
Another quarter-wave plate is inserted in the reference arm in order to 
maximize the visibility of the fringes in the interference pattern, which is 
recorded by a CCD camera (cf. Fig.~\ref{setup}).

\begin{figure}[t!]
\centering
\includegraphics[width=\columnwidth]{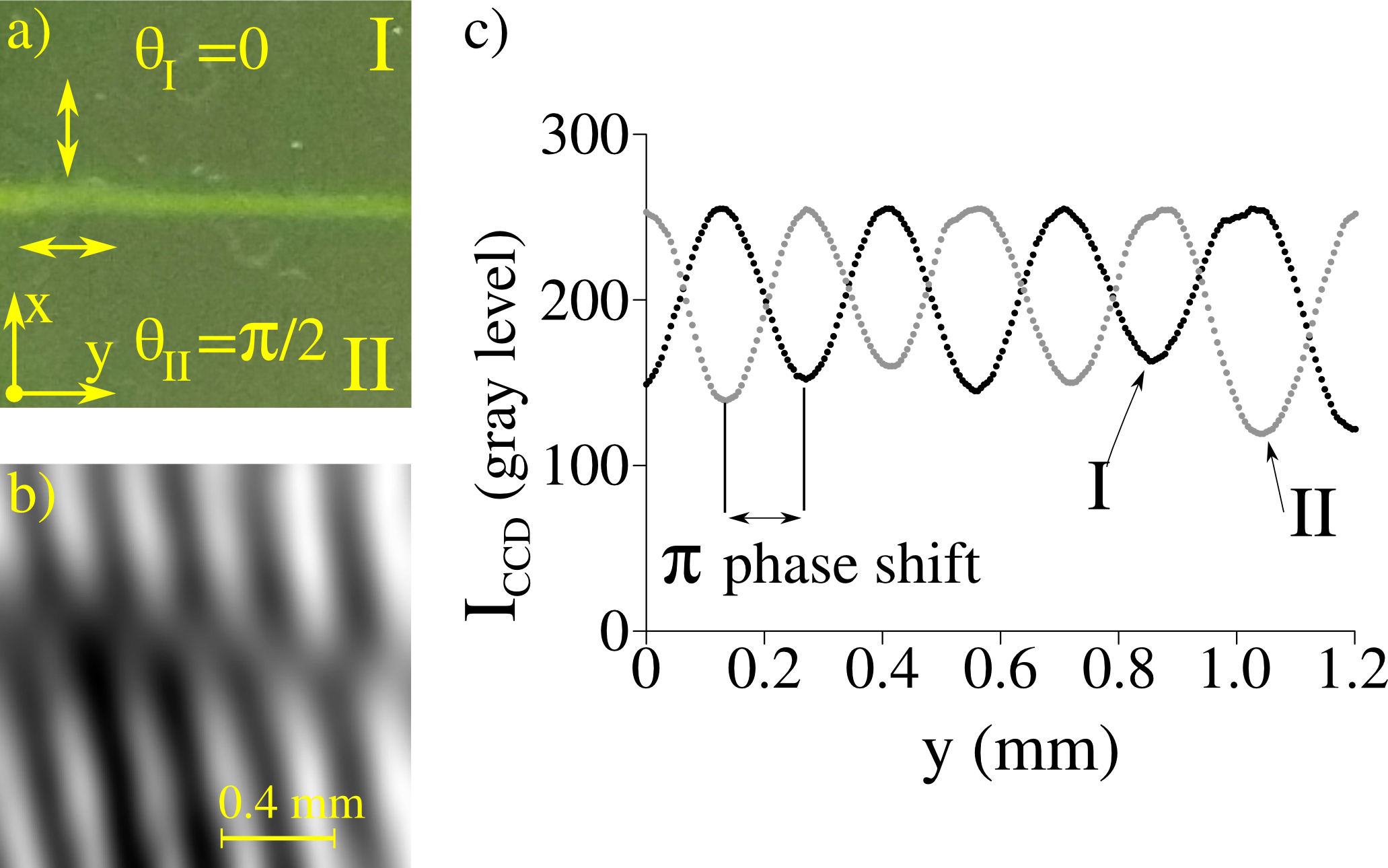}
\caption{(color online). (a) Instantaneous snapshot of the CLC cell showing the interface between regions I and II corresponding to the orthogonal anchoring $\theta_I=0$ and $\theta_{II}=\pi/
2$. (b) Interference pattern recorded for a probe beam hitting the cell across the interface; dislocation lines  evidence the $\pi$ phase shift between regions I and II. (c) One-dimensional 
intensity profiles taken on the interference pattern in regions I and II, respectively.}
\label{phase}
\end{figure}

Figure \ref{phase}a displays an instantaneous  
snapshot of the CLC cell, showing the interface between regions I and II.
An experimentally recorded interference pattern is shown in Fig.~\ref{phase}b. 
Because the probe beam overlaps two regions with different orientation of the 
optical axis, a phase shift is expected to appear in the fringe interference 
pattern. Indeed, dislocation lines can clearly be distinguished across the 
interface, evidencing the $\pi$ phase difference between regions I and II, 
corresponding to $\Phi_{B}=2  \sigma \theta=\pi $, with $\sigma=1$ and 
$\theta=\pi/2$ in this experiment. Figure \ref{phase}c shows two one-dimensional 
intensity profiles taken on the fringe interference pattern in regions I and II, 
respectively. Again, the geometric phase shift can be clearly appreciated.

\section{Conclusion}
In conclusion, we have demonstrated that a Berry phase exists whenever circularly 
polarized light is Bragg-reflected by a CLC layer. Such geometrical phase originates 
from the orientation of the chiral layer at the entrance plane of the cell. 
Using more advanced alignment techniques based on photopolymers 
\cite{Chigrinov_book} and holographic \cite{Escuti_Optica2015,Chigrinov_SID2015} 
recording or direct laser writting \cite{Miskiewicz:OE:2014,Escuti_Optica2015}, 
nearly arbitrary profile of planar anchoring can be achieved. Arbitrary
profiles of geometric phase can then be achieved with application in complex 
wavefront shaping without the use of back reflectors.  A very recent 
application was  demonstrated by Kobayashi et al \cite{Kobayashi:NatPhot:2016}; 
but reference to geometric phase was not  mentioned. Moreover, operation in 
different spectral regions could easily be achieved by changing the reflective 
bandgap of the CLC layer, and broadband devices  could be obtainded 
\cite{Yanming:SPIE:2012} using the multitwisted layer architecture 
\cite{Komanduri:OE:2013}.

\begin{acknowledgments}
R.~Barboza acknowledges FONDECYT POSTDOCTORADO 3140577 for the  financial support. M.G.~Clerc thanks the financial support of FONDECYT 1150507.
\end{acknowledgments}

\bibliography{reference}
\end{document}